\documentclass[aps,prl,twocolumn,groupedaddress]{revtex4-1}
\usepackage{graphicx}
\usepackage{bm}
\usepackage{amsmath}

\begin{document}


\title{Orbital excitation blockade and algorithmic cooling in quantum gases}


\author{Waseem S. Bakr}
\author{Philipp M. Preiss}
\author{M. Eric Tai}
\author{Ruichao Ma}
\author{Jonathan Simon}
\author{Markus Greiner}
\affiliation{Department of Physics, Harvard University, Cambridge,
Massachusetts, 02138, USA}


\date{\today}

\begin{abstract}
Interaction blockade occurs when strong interactions in a confined
    few-body system prevent a particle from occupying an otherwise accessible
    quantum state. Blockade phenomena reveal the underlying granular nature of
    quantum systems and allow the detection and manipulation of the
    constituent particles, whether they are electrons\cite{grabert_single_1992}, spins\cite{ono_current_2002},
    atoms\cite{cheinet_counting_2008,urban_observation_2009,gaetan_observation_2009},
    or photons\cite{birnbaum_photon_2005}. The diverse applications range from single-electron
    transistors based on electronic Coulomb
    blockade\cite{kastner_single-electron_1992} to quantum logic gates
    in Rydberg
    atoms\cite{isenhower_demonstration_2010,wilk_entanglement_2010}. We have
    observed a new kind of interaction blockade
    in transferring ultracold atoms between orbitals in an optical
    lattice. In this system, atoms on the same lattice site undergo coherent
    collisions described by a contact interaction whose strength depends
    strongly on the orbital wavefunctions of the atoms. We induce coherent
    orbital excitations by modulating the lattice depth and observe a
    staircase-type excitation behavior as we cross the interaction-split
    resonances by tuning the modulation frequency. As an application of orbital excitation blockade (OEB), we demonstrate a novel algorithmic route for cooling quantum gases. Our realization of algorithmic cooling\cite{boykin_algorithmic_2002,baugh_experimental_2005} utilizes a sequence of reversible OEB-based quantum operations that isolate the entropy in one part of the system, followed by an irreversible step that removes the entropy from the gas. This work opens the door to cooling quantum gases down to ultralow entropies, with implications for developing a microscopic understanding of strongly correlated electron systems that can be simulated in optical lattices\cite{lewenstein_ultracold_2007,bloch_many-body_2008}. In addition, the close analogy between OEB and dipole blockade in Rydberg atoms provides a roadmap for the implementation of two-qubit gates\cite{schneider_quantum_2011} in a quantum computing architecture with natural scalability.
    \end{abstract}

\pacs{}
\keywords{optical lattices, quantum simulation, bose-hubbard}

\maketitle


An ultracold gas of bosonic atoms in the ground band of an optical lattice is
described by the Bose-Hubbard model\cite{jaksch_cold_1998}, in which atoms can tunnel between
neighboring sites and interact via an onsite repulsive contact interaction. In
a deep lattice where the interactions dominate, the ground state of the system
is a Mott insulator with a fixed atom number per site that is locally constant
over a region of the insulator\cite{greiner_quantum_2002}. The energy per site in the absence of
tunneling is $\frac12 U_\text{gg}n(n-1)$, where $U_\text{gg}$ is the interaction energy for two atoms in
the ground lattice orbital state and $n$ is the atom number on the site. The
Mott state exhibits a transport blockade phenomenon in which the presence
of an atom on a site energetically prevents tunneling of a neighboring atom
onto that site even in the presence of a small bias between the sites. The
transport is blocked unless the bias makes up for the interaction cost,
making it possible, for example, to count atoms tunneling across double-wells
in a superlattice\cite{cheinet_counting_2008}. In this work, we explore an excitation blockade
phenomenon that does not involve transport in the lattice. The excitation
transfers localized atoms between different orbitals on the same site through
modulation of the lattice depth at a frequency close to a vibrational
resonance. Physics in higher optical lattice orbitals has been the focus of
much recent experimental work including the study of dynamics in higher
orbitals\cite{muller_state_2007}, multi-orbital corrections to the interaction
energy\cite{will_time-resolved_2010}, and unconventional forms of
superfluidity involving higher
orbitals\cite{wirth_evidence_2011,soltan-panahi_multi-component_2010}.

\begin{figure}[!h]
    \centering
    \includegraphics[width=84mm]{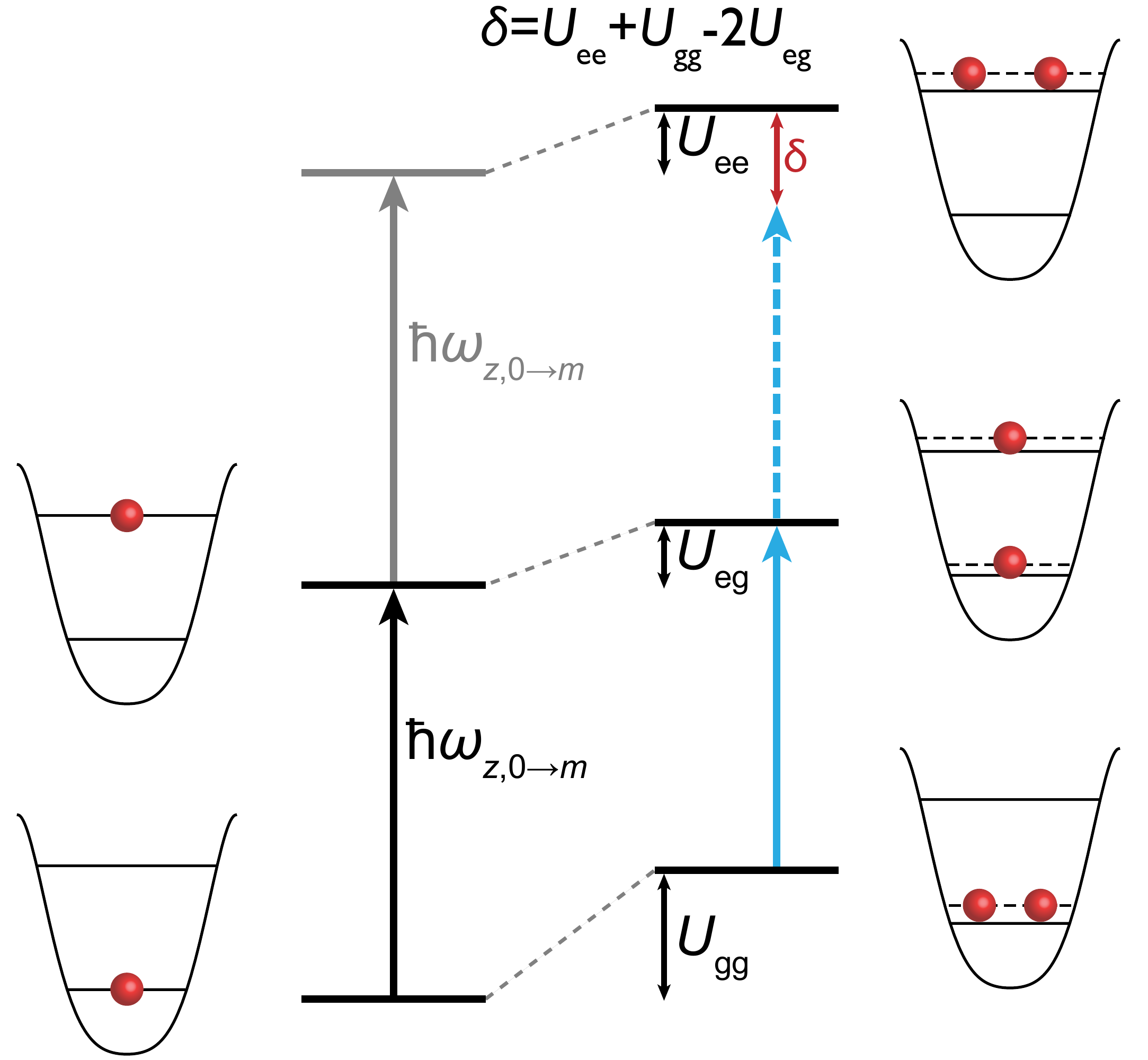}
    \caption{\label{fig1} Orbital excitation blockade mechanism in an optical
    lattice. A single atom on a site is excited to a higher orbital by resonantly
    modulating the lattice depth. For two atoms on the same site, interactions
    lead to an orbital-dependent energy shift. Modulation at the appropriate
    frequency excites one of the atoms to the higher orbital, but is
    off-resonant for exciting the second with a blockade energy $\delta$.}
\end{figure}

The OEB mechanism can be understood in the simplest scenario for two atoms in
a single site of a deep three-dimensional lattice, in which the vibrational
frequencies in all three directions are taken to be different to avoid
degeneracies. The lattice depth along the {$z$-direction} is modulated weakly,
which in the presence of anharmonicity of the lattice potential drives atoms
between the ground orbital and a single, specific excited {$z$-orbital}, subject
to a selection rule that only allows coupling to orbitals of the same
symmetry. For a single atom, excitation to the $m$th orbital requires modulation
at a frequency $\omega_{z,0\to m}$ which is approximately $m\omega_{z,0\to1}$ ignoring the
anharmonicity of the onsite potential. With more than one atom on a site, the
interaction introduces an orbital-dependent shift of the energy levels as
shown in Fig.~\ref{fig1}. In general, the interaction shifts $U_\text{gg}$,
$U_\text{ge}$, and $U_\text{ee}$ are all different and the differences are a
significant fraction of $U_\text{gg}$, where $g$($e$) denotes atoms in the
ground (excited) orbital. If the coupling strength
due to the modulation is small compared to these differences and the
modulation frequency is tuned to $\omega_{z,0\to m} + (U_\text{ge}-
U_\text{gg})/\hbar$, only a single atom is
transferred to the higher orbital and the transfer of a second atom is
off-resonant. In this sense, the first excitation blocks the creation of a
second excitation. 

\begin{figure}
    \centering
    \includegraphics[width=84mm]{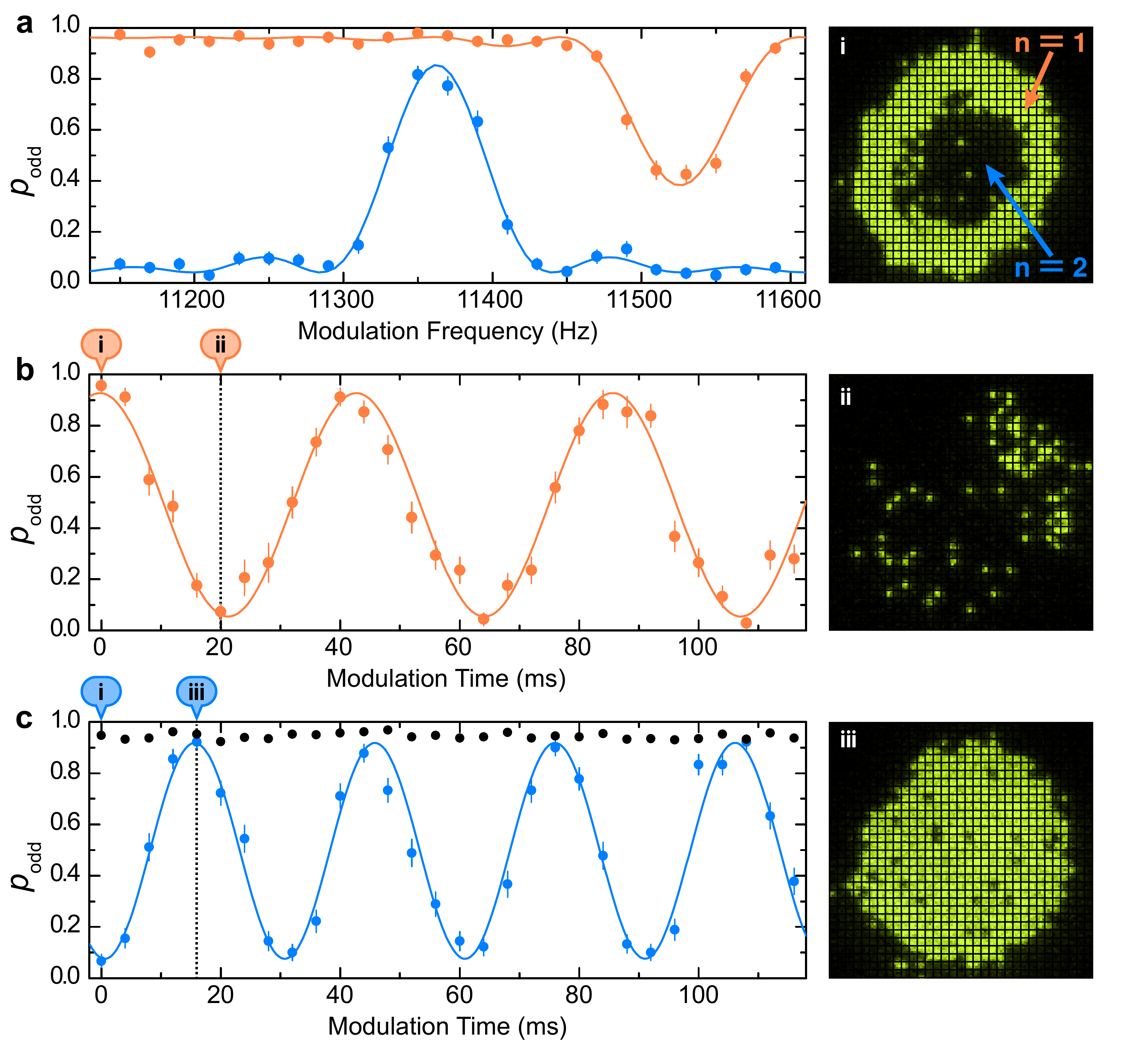}
    \caption{\label{fig2} Time, frequency and site-resolved coherent transfer
    of atoms in a Mott insulator between orbitals. \textbf{a}, Excitations
    transferring a single atom in the $n=1$ (orange) or $n=2$ shell (blue)
    from the ground to the second excited orbital are spectroscopically
    resolved in a two-shell Mott-insulator. \textbf{b}, Rabi oscillations
    between the two orbitals are observed by driving at the resonant
    frequencies for atoms in the $n=1$ shell and \textbf{c}, $n=2$ shell of a
    Mott insulator. Bose-enhancement leads to faster oscillations in the $n=2$
    shell. When the atom number is reduced to obtain one atom per site in the
    region previously containing two atoms, the interaction shift suppresses
    oscillations (black). All error bars are one standard error of the mean.
    (\textbf{i}-\textbf{iii}) Site-resolved snapshots of the Mott insulator
    are shown at different points in the Rabi cycles.}
\end{figure}

The experimental system has been described in previous
work\cite{bakr_probing_2010}. A
two-dimensional Bose-Einstein condensate of rubidium atoms resides in a single
plane of a one-dimensional optical lattice, henceforth referred to as the
axial lattice, with a vibrational frequency of $\omega_{z,0\to1} = 2\pi \times
5.90(2) \, \text{kHz}$. The $z$-axis is perpendicular to the plane and points along the direction of gravity.
In addition, we introduce a lattice in the plane with a spacing of $a=680 \, \text{nm}$ and
a depth of $45E_r$ (trap frequency of $17 \, \text{kHz}$), where $E_r=h^2/8ma^2$ is the recoil energy of the effective lattice wavelength, with $m$ the mass of $^{87}$Rb. The resulting Mott insulator is at the
focus of a high resolution optical imaging system capable of detecting atoms
on individual lattice sites through fluorescence imaging.  Light-assisted
collisions at the start of the imaging process reduce the occupation of a site
to its odd-even parity\cite{bakr_probing_2010}. 

\begin{figure}
    \centering
    \includegraphics[width=84mm]{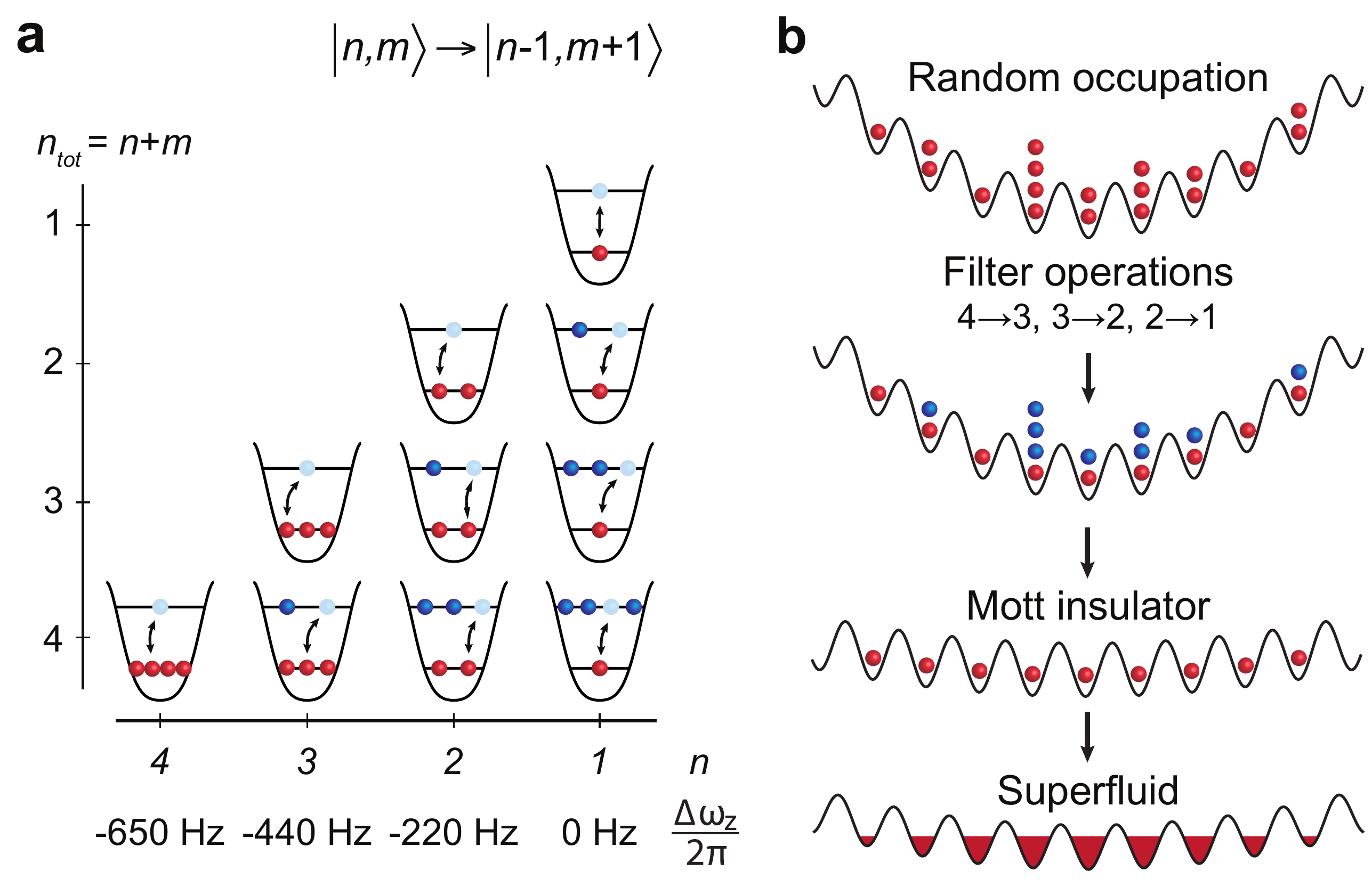}
    \caption{\label{fig3} Algorithmic cooling in an optical lattice.
    \textbf{a}, Landau-Zener chirp for transferring entropy from the ground to
    the fourth band. The lattice modulation frequency is swept across the
    transition resonances from left to right. The interaction shifts $\Delta
    \omega_z$ for excitation of one of $n$ atoms to the fourth orbital
    relative to the excitation frequency for a single atom on a site, are
    shown for different orbital occupations. Excitation processes in the same
    column happen at almost the same frequency to within 30 Hz (see
    Supplementary Table \ref{suptable}). \textbf{b}, A state with random occupation in a
    deep lattice is far from the Mott insulating ground state. Sequential
    filtering operations followed by reduction of the confinement prepares the
    ground state, which can be adiabatically converted to a thermalized
    superfluid in a shallow lattice. Red (blue) spheres denote atoms in the
    ground (excited) band.}
\end{figure}

We start by demonstrating coherent driving of atoms in a Mott insulator
between two orbitals. In the presence of a harmonic trap, the atoms in a 2D
Mott insulator are arranged in concentric rings of fixed atom number per site,
known as shells, with the largest occupation at the
center\cite{bakr_probing_2010}. We prepare a
Mott insulator with two shells and modulate the axial lattice depth by
$\pm1.1(1)\%$ at a frequency chosen to transfer atoms from the ground state to
the second excited orbital. A modulation frequency corresponding to exactly
$\omega_{z,0\to2}$ is resonant for atoms in the outer shell with one atom per
site ($n=1$). Excitation to the fourth excited orbital is suppressed because of an energy
shift of $h \times 1200(80) \, \text{Hz}$ owing to the anharmonicity of the onsite potential.
Rabi oscillations between the states $|g\rangle$ and $|e\rangle$ are detected by lowering the
axial lattice depth at the end of the modulation so that the excited orbital
state becomes unbound and any population in it escapes along the $z$-axis due to
gravity. The Rabi oscillations in that shell, shown in Fig.~\ref{fig2}b, have a
frequency $\Omega = 2\pi \times 23.3(2) \, \text{Hz}$. 

The OEB is demonstrated in the inner $n=2$ shell by modulating at a frequency of
${\omega_{z,0\to2} + (U_\text{ge}- U_\text{gg})/ \hbar}$. For our parameters,
$U_\text{gg}$, $U_\text{ge}$, and $U_\text{ee}$ are ${h \times 480(30)},
360(20), 310(20) \, \text{Hz}$ respectively. The Rabi oscillations between
$|g,g\rangle$ and $1/\sqrt2(|e,g\rangle + |g,e\rangle)$ are detected as an
oscillation of the parity between even
and odd after ejecting the atom in the excited orbital and are shown in
Fig.~\ref{fig2}c. For the same modulation amplitude as before, the oscillations are expected
to occur $\sqrt2$ times faster than the resonant oscillation in the $n=1$ shell owing
to Bose-enhancement\cite{gaetan_observation_2009}. We indeed observe a frequency ratio of 1.42(1)
between the oscillation frequencies. A full frequency spectrum in the two
shells is shown in Fig.~\ref{fig2}a and the frequency separation of the two
resonances of 160(10) Hz matches well with the theoretically expected value of
165(35) Hz when the impact of virtual orbital changing collisions is included
(see Methods).

We next employ OEB to demonstrate a new path to cooling quantum gases.
Evaporative cooling has been the workhorse technique for cooling atomic gases
to nanokelvin temperatures. However, current interest in studying the physics
of strongly-correlated materials, such as high-$T_c$ cuprates, using ultracold
gases\cite{lewenstein_ultracold_2007,bloch_many-body_2008} has spurred
research into developing new cooling techniques that can reach the requisite
picokelvin regime\cite{mckay_cooling_2011,medley_spin_2011}. The field of
quantum information offers an alternative cooling paradigm, wherein a sequence
of unitary quantum gates purifies a subset of the qubits in a system by moving
entropy and isolating it in another part of the
system\cite{boykin_algorithmic_2002}. One realization of such a cooling
scheme, heat-bath algorithmic cooling, has been successfully demonstrated with
solid-state nuclear magnetic resonance qubits\cite{baugh_experimental_2005}.
We introduce an analogous technique for quantum gases where the unitary
operations are achieved using OEB, building on previous theoretical proposals
in this
direction\cite{rabl_defect-suppressed_2003,popp_ground_2006,nikolopoulos_atom-number_2010,sherson_shaking_2010}.

\begin{figure*}
    \centering
    \includegraphics[width=172mm]{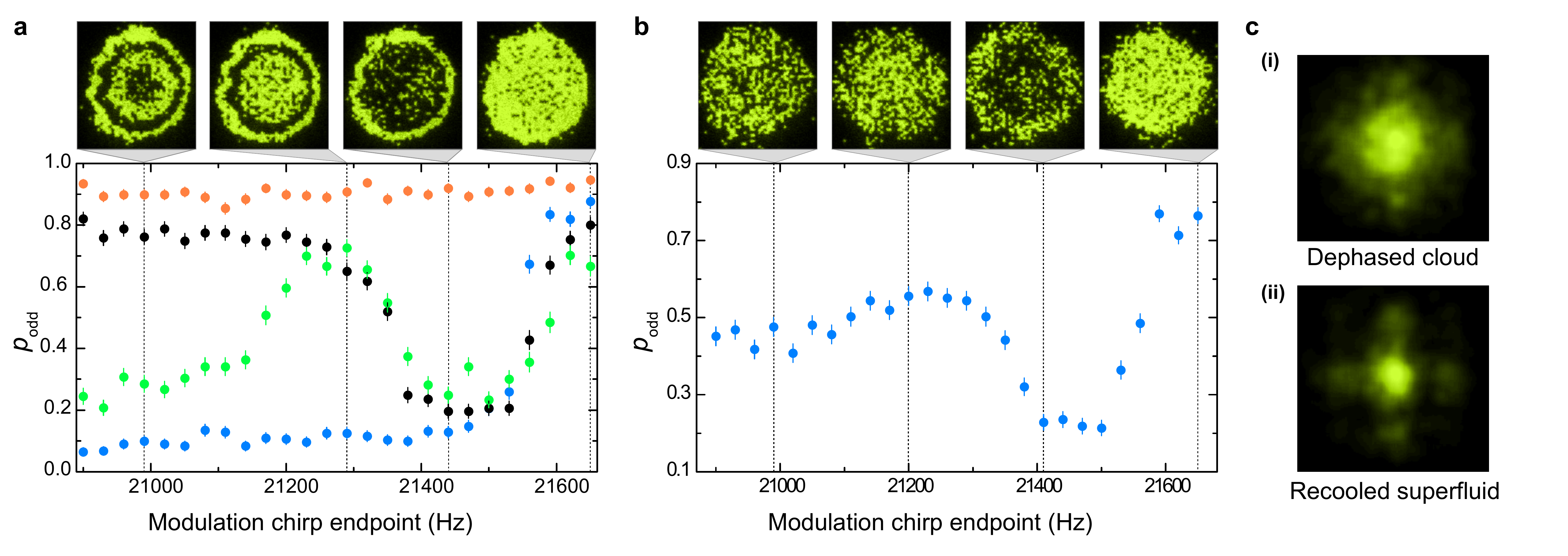}
    \caption{\label{fig4} Experimental realization of algorithmic cooling.
    \textbf{a}, By chirping the modulation towards higher frequencies, atoms
    in a four-shell Mott insulator are sequentially excited one at a time to
    the fourth orbital. Population in the higher orbital is subsequently
    ejected at the end of the chirp. The average parity in the $n=1$ (orange),
    $n=2$ (blue), $n=3$ (black) and $n=4$ (green) shell is shown at different
    points in the chirp, together with single shot images, illustrating the
    conversion to a three, two and finally one shell insulator. \textbf{b},
    The same frequency chirp algorithmically cools a state with random
    occupancy into an $n=1$ Mott insulator, observed as an enhancement in odd
    occupancy. All error bars are one standard error of the mean. \textbf{c},
    (\textbf{i}) An incoherent cloud does not exhibit an interference pattern
    in a 5 ms time of flight expansion after adiabatically lowering the
    lattice depth. (\textbf{ii}) Cooling converts the incoherent cloud to a
    Mott insulator in the deep lattice. After adiabatically lowering the
    lattice depth, a superfluid forms and an interference pattern is obtained
    in the expansion images.}
\end{figure*}

A bosonic quantum gas at a finite temperature $T$ adiabatically loaded into
the ground band of a optical lattice stores its entropy in the form of atom
number fluctuations in the zero-tunneling limit. Within the local density
approximation, a lattice site with a local chemical potential $\mu$ is
described by a density matrix $\hat{\rho}=\sum_n p_n |n\rangle\langle n|$,
where $p_n$, the probability of having $n$ atoms on the site, is given by
$e^{-\beta \left(\frac{1}{2}U_{gg}n(n-1)-\mu n\right)}/Z$. Here, $\beta=1/k_B
T$ and $Z$ is the grand canonical partition function. Cooling to zero
temperature is achieved by changing the atom number distribution on each site
to obtain $p_n=\delta_{n,\lceil\mu/U_{gg}\rceil}$.

The crucial ingredient for algorithmic cooling is a unitary operation that
realizes the transformation $|n,m\rangle \rightarrow |n-1,m+1\rangle$ for each
$n$ separately, where $|n,m\rangle$ denotes a Fock state with $n$ atoms in the
ground band and $m$ atoms in an excited band. Resonant lattice modulation in
the presence of OEB results in a rotation gate
$\hat{R}_{nm}(t)=\exp{[i(\Omega_{nm}t|n-1,m+1\rangle\langle n,m|+c.c.)]}$,
where $\Omega_{nm}$ is the transition's Rabi frequency and the required
transformation is achieved for a modulation time ${t = \pi/2\Omega_{nm}}$.
Entropy is transferred from the ground band to the excited band by performing
a sequence of $\pi$ gates $\hat{R}_{N-s,s}$ from $s=0$ to $s=N-1$ with $N$
chosen large enough such that $p_N\approx 0$ in the initial state. At the end
of this sequence, most of the entropy of the gas has been transferred to the
excited band and can be removed from the system by ejecting the atoms in that
band. The local chemical potential $\mu$ is readjusted to recover a situation
closer to thermal equilibrium by reducing the harmonic confinement to a new
value $\omega_{\textrm{low}}$ so that $\mu<U_{gg}$ throughout the gas. At this
point, residual entropy is stored in the resulting $n=1$ Mott insulator in the
form of holes that are preferably located near the edge of the cloud. The gas
is allowed to rethermalize by lowering the lattice depth to allow tunneling,
and the final entropy of the thermalized state would be significantly reduced
compared to the initial state.

We start by experimentally demonstrating the algorithm on a state with known
atom number, namely a four shell Mott insulator, and reducing the site
occupation everywhere in the insulator to a single atom per site. To increase
the blockade energy for this set of experiments, we transfer atoms to the
fourth axial orbital rather than the second. We also replace the rotation
gates demonstrated in the first part of this work with Landau-Zener
transitions to improve the fidelity of the algorithm. We linearly sweep the
modulation frequency from 20.90 kHz to 21.65 kHz in 250 ms. The chirp realizes
a sequence of quantum operations that transfer atoms to the excited orbital
one at a time, until only one atom remains in the ground state in all shells
(Fig.~\ref{fig3}a). We probe the ground orbital occupancy at different points
in the frequency chirp by ejecting atoms in the higher orbital as before and
then performing the parity imaging. The parity of the different shells during
the chirp is shown in Fig.~\ref{fig4}a, and an analogue of the typical Coulomb
blockade staircase is seen in the data. Shell-sensitive manipulation of a Mott
insulator had been achieved in previous experiments using a microwave
transition between hyperfine states, but the lack of a strong blockade allowed
transfer of only a small fraction of the population to the target
state\cite{campbell_imaging_2006}. 

Next, we demonstrate cooling by performing the algorithm on a state that is
far from the many-body ground state. To prepare such a state, we
non-adiabatically load a condensate into a deep lattice, projecting the
wavefunction onto a state with Poissonian site occupancy that rapidly loses
coherence between sites. Using the same operation sequence as before, we
progressively reduce the randomness of the ground band occupancy, preparing a
single-occupancy Mott insulator (Fig.~\ref{fig3}b). We enhance the odd
occupancy from 0.45(1) to 0.76(2) (Fig.~\ref{fig4}b), demonstrating
significant atom number squeezing limited by the conversion efficiency of the
Landau-Zener transitions. To complete the algorithm, we readjust the harmonic
confinement to obtain a state close to the many-body ground state. We verify
the ground state character by ramping back adiabatically to a $5E_r$ lattice
in 100 ms and releasing the atoms from the lattice. Without cooling, we obtain
a featureless cloud shown in Fig.~\ref{fig4}c(i), indicating an absence of the
coherence expected in the superfluid ground state. With cooling, the Mott
insulator is adiabatically converted to a superfluid, giving rise to matter
wave interference peaks shown in Fig.~\ref{fig4}c(ii).

We now discuss the limits on the entropies that can be achieved with
algorithmic cooling. The conversion efficiency to a single-occupancy Mott
insulator is technically limited in our system by heating due to spontaneous
emission during the sweep and to a lesser extent, by the efficiency of the
Landau-Zener sweep for clouds with large average occupancies (see Methods).
While we have demonstrated cooling of hot clouds, the single-occupancy
probability we have achieved using algorithmic cooling in a two-shell Mott
insulator is 0.94(1). This is comparable to what had been previously achieved
with evaporative cooling, corresponding to an average entropy of $0.27 k_B$
per particle~\cite{bakr_probing_2010}. Nevertheless, lattice heating can be
made negligible by using a further-detuned lattice (e.g. 1064~nm), while
shaped pulses can improve the Landau-Zener transfer
efficiency\cite{popp_ground_2006}. More fundamentally, the single-shot cooling
algorithm we have implemented is limited by initial holes in the Mott
insulator which cannot be corrected. However, repeated iterations of the
algorithm can circumvent this problem and bring the cloud to zero entropy with
quick convergence\cite{popp_ground_2006}. The cycle alternates between (a)
using OEB to produce a reduced entropy $n=1$ insulator in a harmonic
confinement $\omega_{\textrm{low}}$ (demonstrated above) and (b) adiabatically
increasing the confinement to $\omega_{\textrm{high}}$ in the presence of
tunneling to move hot particles to the center of the cloud where they can be
removed again. Alternatively, the outer edge of the cloud containing the holes
can be removed using the high resolution available in our
system\cite{weitenberg_single-spin_2011}.

In conclusion, we have observed a new blockade phenomenon in optical lattices
when exciting atoms to higher orbitals, analogous to dipole excitation
blockade in Rydberg atoms. The blockade permits deterministic manipulation of atom number in an optical lattice. We have used it to convert a multi-shell Mott insulator into a singly-occupied insulator with over a thousand sites, the largest quantum register achieved so far in an addressable system. The same technique allows initialization of registers in longer wavelength lattices where a Mott insulator cannot be prepared~\cite{weiss_imaging_2007}. 
OEB also opens a route to implementing
quantum gates in optical lattices. Single-site
addressing\cite{weitenberg_single-spin_2011}, possible with our microscope,
can perform rotations of individual orbital-encoded qubits rather than the global rotations demonstrated in this work. Controlled-NOT gates can be
implemented by conditionally moving the control qubit onto the target qubit
site, and performing an interaction-sensitive rotation of the target
qubit\cite{schneider_quantum_2011}. Finally, the algorithmic cooling technique we have developed could potentially achieve the ultralow entropies required for quantum simulation~\cite{lewenstein_ultracold_2007,bloch_many-body_2008} and computation in optical lattices, and establishes a bridge to quantum information for importing novel ideas for cooling quantum gases.

\begin{acknowledgments}
We would like to thank S. F\"olling for stimulating discussions. This work was
supported by grants from the Army Research Office with funding from the DARPA
OLE program, an AFOSR MURI program, and by grants from the NSF.
\end{acknowledgments}

\bibliography{oeb_paper}

\section*{Methods}
\setcounter{figure}{0}
\makeatletter 
    \renewcommand{\thefigure}{S\@arabic\c@figure} 
\makeatother 

\subsection{State Preparation}
Our experiments begin with a degenerate 2D Bose gas of $^{87}$Rb atoms
prepared in the $|F=1,m_f=-1\rangle$ state in a single layer of a 1D optical
lattice with spacing 1.5 $\mu$m, in the focal plane of a high resolution
imaging system as described in previous work. The atoms are then loaded into a
2D optical lattice with spacing 680 nm, which is ramped up to a depth of
$45E_r$ adiabatically on either a single- or many- body timescale, depending
upon the experiment to be performed.

\subsection{Higher Band Removal}
The orbital blockade is observed through the deterministic removal of atoms in
higher bands of the 1.5 $\mu$m lattice. For removal of atoms from the second band, this is achieved by reducing the depth of this
lattice to 3.8 kHz from an initial depth of 35.8 kHz. Gravity produces a shift of 3.2 kHz per well, thus
inducing second band atoms to Zener tunnel away within a few ms. The Landau-Zener tunneling rate from the ground state is given by $\Gamma_{LZ}=\frac{mga}{2\pi\hbar}e^{-g_c/g}\approx 12$ Hz. Here $g$ is the gravitational acceleration and $g_c=a\omega_z^2/4$. This effect leads to a loss of ground state atoms on the percent level, but can be made negligible by using excitations to the fourth band.

\subsection{Band Dependent Energy Shifts} 
Due to its large spacing, the 1.5 $\mu$m lattice has a recoil energy of only 250
Hz. Consequently its depth is $\sim150$ recoils, and its low-lying eigenstates
are, to good approximation, the Hermite-Gaussians of a harmonic oscillator.
The interaction energy between particles in bands $m$ and $n$ may thus be written in terms of the ground band
interaction energy $U_{00}$ as:
\[U_\text{\textit{nm}} = U_{00} (2- \delta_{nm}) \frac{\int |\psi_m(x)|^2 |\psi_n(x)|^2 \, dx}{\int |\psi_0(x)|^4 \, dx}\]
where $\psi_m(x)$ is the normalized $m$th harmonic oscillator wavefunction. The total
interaction shift for $M$ particles in band $m$, and $N$ particles in band $n$ is
thus:
\[\frac{M(M-1)}{2} U_\text{\textit{mm}} + \frac{N(N-1)}{2} U_\text{\textit{nn}} +
MNU_\text{\textit{mn}}.\]
The interactions also produce off-resonant band changing collisions, with a Rabi frequency:
\[\Omega_{mn \leftrightarrow pq} = U_{00} C_{mn \leftrightarrow pq} \frac{\int
\psi_m(x) \psi_n(x) \psi_p^\ast(x) \psi_q^\ast(x) \, dx}{\int|\psi_0(x)|^4 \,
dx}\]
up to a combinatoric factor $C_{mn \leftrightarrow pq}$ resulting from Bose
enhancement. For an energy defect of ${\delta_{mn\leftrightarrow pq} \gg
\Omega_{mn\leftrightarrow pq}}$, this process produces an energy shift of
$\Delta_\text{\textit{mn}} = -|\Omega_{mn \leftrightarrow pq}|^2 / \delta_{mn
\leftrightarrow pq}$.
For our experiment, the dominant band changing collision is $|m=0,n=2\rangle
\to |p=1,q=1\rangle$ with a Rabi frequency $\Omega_{02 \leftrightarrow 11} =
2\pi \times 120 \, \text{Hz}$. In this case lattice anharmonicity and
interaction shifts produce an energy defect of  $\delta_{02 \leftrightarrow11}
= 2\pi \times 330 \, \text{Hz}$, resulting in an additional overall shift of
the $|0,2\rangle$ state of $\Delta_{02} = -2\pi \times 45 \, \text{Hz}$.

\begin{table}
\begin{tabular}{ccccc}
	\hline

	$\Delta\omega_{z}/2\pi\,[\text{Hz}]$ & $n=1$ & $n=2$ & $n=3$ & $n=4$\\

	\hline\hline
	$m=0$ & 0    & -217 & -434 & -651 \\
	$m=1$ & -9   & -226 & -443 &      \\
	$m=2$ & -18  & -235 &      &      \\
	$m=3$ & -27  &      &      &      \\
	\hline
\end{tabular}
\caption{\label{suptable}Frequency shifts in Hz for a transition transferring an atom from the ground
band to the fourth band, starting with $n$ and $m$ atoms in each of these
bands. The shifts are measured relative to an initial state with one atom in
the ground band and none in the excited band.}
\end{table}

\subsection{Limits on entropies achievable with algorithmic cooling} 
In our system, the main limitation on the achievable entropies with
algorithmic cooling is losses during the Landau-Zener chirp. Spontaneous
emission after absorption of photons from the lattice leads to excitation of
atoms from the ground state at a rate of $0.1\,\text s^{-1}$. These atoms are quickly
lost due to tunneling in the higher bands in the case of the in-plane lattice
or in the band filtering step in the case of the axial lattice. This leads to
holes in an $n=1$ Mott insulator (3\% during the 250 ms ramp), setting a lower
bound on the reachable entropy after thermalization of $\approx 0.15 k_B$ per
particle. The lattice light is detuned 25 nm to the blue of the atomic
resonance, and the heating rate can be made negligible by increasing the
detuning (e.g. using 1064 nm lattice light). A more fundamental limit on the
single-shot cooling algorithm demonstrated here is given by initial holes in
the Mott insulator that cannot be corrected. For our Mott insulators, the hole
fraction is on the order of 0.5\% corresponding to a post-thermalization
entropy of $\approx 0.06 k_B$ per particle. This limit can be overcome by the
iterative algorithm described in the text. For Mott insulators with large
initial atom numbers per site, it is also important to take into account the
efficiency of the Landau-Zener chirp. For excitation to the second band, the
measured efficiency is 0.94(1). 





%
\end{document}